\documentclass[12pt]{article}
\usepackage{epsfig}
\newcommand{\be}{\begin{equation}}
\newcommand{\ee}{\end{equation}}

\title{\Large\bf Orbitally excited $D$ and $B$ mesons in the approach of the QCD string with quarks at the ends}
\author{\large Yu.S.Kalashnikova\thanks{E-mail: yulia@heron.itep.ru}\hspace*{2mm}$^{\rm a}$,
A.V.Nefediev\thanks{E-mail: nefediev@heron.itep.ru
nefediev@cfif.ist.utl.pt}\hspace*{2mm}$^{\rm a,b}$\\[5mm]
${\rm ^a}$ {\normalsize\it Institute of Theoretical and Experimental Physics,}\\
{\normalsize\it 117218, B.Cheremushkinskaya 25, Moscow, Russia}\\
${\rm ^b}$ {\normalsize\it Grupo Te\'orico de Altas Energias (GTAE),}\\
{\normalsize\it Centro de F\'\i sica das Interac\c c\~oes Fundamentais (CFIF),}\\
{\normalsize\it Departamento de F\'\i sica, Instituto Superior T\'ecnico,} \\
{\normalsize\it Av. Rovisco Pais, P-1049-001 Lisboa, Portugal}}
\date{}         
\begin{document}
\maketitle

\begin{abstract}
In this letter we discuss the masses and the splittings of $1^{2S+1}P_J$ states in the 
spectrum of $D$ and $B$ mesons,
as they appear in the approach of the QCD string with quarks at the ends. We find good
agreement of our predictions with those of other QCD-motivated models as well as with 
the lattice and experimental data, including recent experimental results.
We discuss the ordering pattern for $P$ levels in $D$- and $B$-mesonic spectrum.
\end{abstract}

Data on the spectroscopy of heavy-light mesons coming from various experimental
collaborations are challenge for theorists, and these are $D$ and $B$ mesons to play an
important role in checks of the validity and accuracy of the models.

In this letter we address questions concerning the masses of orbitally excited $D$ and 
$B$ mesons in the method of the QCD string with quarks at the ends,
paying special attention to the ordering of the $P$ levels. The choice of
$D$ and $B$ mesons is not accidental and is stipulated by the recent data
on the masses and decays of the above mentioned heavy-light mesons
coming from various experimental collaborations. Despite of the fact that there is no agreement between
them and some resonances are not yet confirmed, still we find it interesting to compare
these experimental data, as well as those provided by other models and lattice
simulations, with the predictions of our approach. First, let us remind the reader
the basic ideas of the latter.

\begin{table}[t]
\begin{center}
\begin{tabular}{|c|c|c|c|c|c|}
\hline
Meson&$\sigma$, $GeV^2$&$\alpha_s$&$C_0$, $MeV$&$m_Q$, $MeV$&$m_q$, $MeV$\\
\hline
$D$&0.17&0.4&196&1400&9\\
\hline
$B$&0.17&0.39&169&4800&5\\
\hline
\end{tabular}
\end{center}
\caption{Parameters of the Hamiltonian (\ref{Hdec})-(\ref{Vsd2}).}
\end{table}

Starting from the gauge-invariant wave function of the $q\bar q$ meson,
\be
\Psi_{q\bar q}(x,y|A)=\bar{\Psi}_{\bar q}(x)\Phi(x,y)\Psi_q(y),
\ee
with $\Phi$ being the parallel transporter, we write the Green's function of the meson,
\be
G_{q\bar q}=\langle\Psi^+_{q\bar q}(\bar{x},\bar{y}|A)\Psi_{q\bar
q}(x,y|A)\rangle_{q\bar{q}A},
\label{Gqq}
\ee
and perform the integration over the quark and the gluonic fields. For the latter case we make
use of the minimal area law asymptotic for the Wilson loop bounded by the quark and the
antiquark trajectories (see, {\it e.g.}, \cite{VCM}),
\be
\left\langle TrP\exp{\left(ig\oint_Cdz_{\mu}A_{\mu}\right)}\right\rangle_A\sim \exp{(-\sigma S_{\rm
min})},
\ee
where $\sigma$ is the string tension in the fundamental representation of the SU(3) colour
group, and the area $S_{\rm min}$ can be approximated by means of the straight-line anzatz
\cite{DKS},
\begin{eqnarray}
S_{\rm min}=\int_0^Tdt\int_0^1d\beta\sqrt{(\dot{w}w')^2-\dot{w}^2w'^2},\nonumber\\
\label{Smin}\\
w_{\mu}(t,\beta)=\beta x_{1\mu}(t)+(1-\beta)x_{2\mu}(t),\nonumber
\end{eqnarray}
with $x_{1,2}$ being the coordinates of the quark and the antiquark. Now, applying the
Feynman-Schwinger representation to the single-quark propagators and introducing the
einbein fields $\mu_{1,2}$ to simplify the relativistic kinematics \cite{einbein}, we,
finally, arrive at the
following expression for the Hamiltonian of the meson \cite{Lisbon,KNS}:
\be
H=H_0+V_{str}+V_{sd},
\label{Hdec}
\ee
\be
H_0=\sum_{i=1}^2\left(\frac{\vec{p}^2+m_i^2}{2\mu_i}+\frac{\mu_i}{2}\right)+\sigma
r-\frac{\kappa}{r}-C_0,\label{H0}
\ee
\be
V_{str}\approx -\frac{\sigma
(\mu_1^2+\mu_2^2-\mu_1\mu_2)}{6\mu_1^2\mu_2^2}\frac{\vec{L}^2}{r}
+\frac{\sigma^2(\mu_1+\mu_2)(4\mu_1^2-7\mu_1\mu_2+4\mu_2^2)}{72\mu_1^3\mu_2^3}\vec{L}^2,\nonumber\\
\label{Vstr}
\ee
$$
V_{sd}=\frac{8\pi\kappa}{3\mu_1\mu_2}(\vec{S}_1\vec{S}_2)
\left|\psi(0)\right|^2-\frac{\sigma}{2r}\left(\frac{\vec{S}_1\vec{L}}{\mu_1^2}+
\frac{\vec{S}_2\vec{L}}{\mu_2^2}\right)
+\frac{\kappa}{r^3}\left(\frac{1}{2\mu_1}+\frac{1}{\mu_2}\right)\frac{\vec{S}_1\vec{L}}{\mu_1}
+\frac{\kappa}{r^3}\left(\frac{1}{2\mu_2}+\frac{1}{\mu_1}\right)
\frac{\vec{S}_2\vec{L}}{\mu_2}
$$
\be
+\frac{\kappa}{\mu_1\mu_2r^3}\left(3(\vec{S}_1\vec{n})
(\vec{S}_2\vec{n})-(\vec{S}_1\vec{S}_2)\right)\label{Vsd2}
+V_{loop}(\kappa^2),
\ee
where in (\ref{Hdec}) we supply the purely nonperturbative interaction, coming from the string-like
picture of confinement, by the perturbative Coulomb interaction $(\kappa=\frac43\alpha_s)$, as well as by the
constant negative shift, $C_0$, due to the light-quark self-energy \cite{Simse} strongly needed
to bring the Regge trajectory intercepts into their experimental values. The term $V_{str}$
deserves special attention, since it is originated from the square root in (\ref{Smin})
and describes the contribution of the QCD string into the total inertia of the rotating
$q\bar q$ system. This contribution is important to establish the correct slope of the
mesonic Regge trajectories \cite{MNS}. We keep the first two terms in its expansion in
powers of $\sigma/\mu^2$. The term $V_{sd}$ contains spin-dependent
interaction generated by both, perturbative and nonperturbative, potentials. Finally, the last term,
$V_{loop}(\kappa^2)$, comes from the one-loop corrections to the potential. It is given by
equations (3.1) and (3.2) of the paper \cite{alpha2} with the obvious change
$m_{1,2}\to\mu_{1,2}$, and we choose the renormalization scale to be equal to the reduced
effective mass $\mu$.
Finally, to fix the Hamiltonian (\ref{Hdec})-(\ref{Vsd2}), we use the values of the parameters listed in Table I.

\begin{table}[t]
\begin{center}
\begin{tabular}{|c|c|c|c|c|c|c|c|c|}
\hline
Meson&$\mu_1$&$\mu_2$&$\mu$&$M_0$&$\Delta E_{str}$&$E_{st}$&$E_{so_1}$&$E_{so_2}$\\
\hline
$D$&1522&597&429&2444&-26&14&15&-13\\
\hline
$B$&4847&675&593&5780&-26&6&7&-10\\
\hline
\end{tabular}
\end{center}
\caption{Solutions of the eigenvalue problem for the Hamiltonian (\ref{H0}) and the
coefficients from equation (\ref{EE}) for the set of parameters given in Table I. All
quantities are given in $MeV$.}
\end{table}

\begin{table}[t]
\begin{center}
\begin{tabular}{|c|c|c|c|c|}
\hline
&$^3P_0$&$^1P_1$&$^3P_1$&$^3P_2$\\
\hline
$^3P_0$&-1,\quad -1,\quad $-1/4$&&&\\
\hline
$^1P_1$&&0,\quad 0,\quad $-1/4$&$1/\sqrt{2},\quad -1/\sqrt{2},\quad 0$&\\
\hline
$^3P_1$&&$1/\sqrt{2},\quad -1/\sqrt{2},\quad 0$&$-1/2,\quad -1/2,\quad 1/4$&\\
\hline
$^3P_2$&&&&$1/2,\quad 1/2,\quad 1/20$\\
\hline
\end{tabular}
\end{center}
\caption{The matrix elements of the spin-dependent ope\-ra\-tors between $P$ states given in the
form: $\langle \vec{S}_1\vec{L}\rangle$, $\langle \vec{S}_2\vec{L}\rangle$,
$\langle (\vec{S}_1\vec{n})(\vec{S}_2\vec{n})\rangle$.}
\end{table}

Einbein fields $\mu_{1,2}$ are kept as variational parameters and the spectrum is
minimized then with respect to them. The extremal values of $\mu$'s play the role of the constituent
masses of the quarks and appear dynamically due to the interaction. This feature of the
given approach allows one to start from the current mass of the constituent
(gluons also can be described in this formalism) and to arrive at its effective
constituent mass self-consistently. However this simple interpretation should be considered 
with caution. The first source of error is neglecting the quark Zitterbewegung. Indeed, 
we neglect the negative-signed solutions for $\mu_{1,2}$ expecting their small influence on 
the spectrum \cite{KNS}. On the other hand, the simple quantum mechanical reduction of the
relativistic field-theory problem given by the QCD string approach is not applicable for
the description of chiral effects, such as the Bogoliubov-type transformation from bare to
dressed quarks and the formation of a nontrivial chiral condensate. Therefore one cannot
pretend to describe the pion in this framework. In realistic quantum-field-theory-based models each mesonic
state possesses two wave functions which describe the motion forward and backward in time
of the $q\bar q$ pair inside the meson \cite{twowave}. For the pion, which is expected to be strictly
massless in the chiral limit, the two wave functions are of the same order of
magnitude, so that none of them can be neglected. Luckily the backward motion is
suppressed if at least one of the quarks is heavy \cite{twowave}, so that one expects to arrive at
reliable predictions in the case of heavy-light mesons. 

Since the Hamiltonian $H_0$, which plays the role of the zeroth order approximation for
the problem, conserves the angular momentum $\vec{L}$, the total spin $\vec{S}$,
and the total momentum $\vec{J}=\vec{L}+\vec{S}$ separately, then its eigenstates can be
specified as terms, $n^{2S+1}L_J$, with $n$ being the radial quantum number. In the remainder of this letter
we shall concentrate on the states with $n=L=1$. Their masses can be represented
as
\be
M(1^{2S+1}P_J)=\langle 1^{2S+1}P_J|H|1^{2S+1}P_J\rangle\hspace*{5cm}
\label{EE}
\ee
$$
\hspace*{3cm}=M_0+\Delta E_{str}+E_{so_1}\langle \vec{S}_1\vec{L}\rangle+E_{so_2}\langle \vec{S}_2\vec{L}\rangle
+E_{st}\langle 3(\vec{S}_1\vec{n})(\vec{S}_2\vec{n})-(\vec{S}_1\vec{S}_2)\rangle,
$$
where $\Delta E_{str}$ is the contribution of the string correction and the term with the
spin-spin interaction does not contribute since the wave function at the origin vanishes
for orbitally excited states, whereas the corresponding one-loop contribution is negligible.
 The results of numerical calculations, including the values
of the coefficients entering equation (\ref{EE}), are listed in Table II (see \cite{KNS} for the
details of the calculations).

In Table III we give the matrix elements of the spin-tensor and spin-orbit operators
between $P$-level states. Since the spin-orbit interaction mixes states with different
total spin, then the masses of the physical states with the total momentum $J=1$ are
subject to a matrix equation,
\be
\left|
\begin{array}{lr}
\langle 1^1P_1|H|1^1P_1\rangle-E&\langle 1^1P_1|H|1^3P_1\rangle\\
\langle 1^3P_1|H|1^1P_1\rangle&\langle 1^3P_1|H|1^3P_1\rangle-E
\end{array}
\right|=0.
\ee

In Tables IV,V we give our predictions for the masses of the $P$-level $D$ and $B$ mesons 
and compare them with the predictions
of other models as well as with the lattice and experimental data coming from
various collaborations.

From Tables IV,V one can deduce several conclusions. First, all three mentioned models give
good description of the $1P_2$ states, whereas all of them fail to reproduce a very
heavy $1P_0$ $B$-mesonic state reported by OPAL \cite{OPAL}. If this
experimental value is 
confirmed, then this
will serve as a signal that all theoretical approaches miss something, and this question
deserves additional careful study. On the other hand, lattice simulations give the mass
$5.754GeV$ for this state \cite{lattice2}, which is also about $100MeV$ lower than the OPAL
value. This stresses once again that the experimental situation strongly needs
clarification. A similar state in the spectrum of $D$ mesons is not reported yet by experimental 
collaborations, though all models and the lattice simulations give a consistent prediction
for it to be around $2430\div 2440MeV$.

Another conclusion which one can make from Tables IV,V is that there is no agreement
concerning $1P_1$ states. Different models give different splitting patterns (see also the
discussion in \cite{lattice2}). 
To have a better insight into the nature of this
splitting let us study the heavy-quark limit, $m_Q=m_1\to\infty$, analytically, 
which is
possible in our approach. Only the coefficient $E_{so_2}$, in notations of equation (\ref{EE}),
survives in this limit, and the expression for it reads
\be
\label{Eso2}
E_{so_2}=-\frac{\sigma}{2\mu^2}\langle r^{-1}\rangle+
\frac{\kappa}{2\mu^2}\langle r^{-3}\rangle
+\frac{9\kappa^2}{16\pi\mu^2}\left[
\left(\frac{19}{18}+\gamma_E\right)\langle r^{-3}\rangle+\langle r^{-3}\ln(\mu r)\rangle
\right],
\ee
where $\gamma_E=0.5772$ is the Euler constant and the averaging is performed over the
zeroth-order wave function $\psi_{nl}(r)$ corresponding to both states,
$P_{1/2}$ and $P_{3/2}$, which are now the true eigenstates of the
Hamiltonian \footnote{We follow the standard notations using the total momentum of the
light quark, $\vec{j}_2=\vec{L}+\vec{S}_2$, as the subscript.}.
As discussed in \cite{yuaDB,KNS}, solution of the eigenstate problem for the Hamiltonian
(\ref{H0}) in this limit is given by solutions to the Schr{\" o}dinger equation
\be
\left(-\frac{d^2}{d\vec{x}^2}+|\vec{x}|-\frac{\lambda}{|\vec{x}|}\right)
\chi_{\lambda}=a(\lambda)\chi_{\lambda},
\label{Schr}
\ee
with the reduced Coulomb-potential strength $\lambda$ being the solution of the equation
(we put the light-quark current mass equal to zero for simplicity)
\be
\lambda^2=\frac43\kappa^2\left(a+2\lambda\left|\frac{\partial
a}{\partial\lambda}\right|\right),
\label{lam0}
\ee
which is $\lambda_0=1.215$ for $\alpha_s=0.39$ and $\lambda_0=1.250$ for $\alpha_s=0.4$. The reduced effective mass
$\mu$ takes the value 
\be
\mu=\frac12\sqrt{\sigma}\left(\frac{\lambda_0}{\kappa}\right)^{3/2}\approx
0.7GeV
\ee
and
\be
\langle r^N\rangle=(2\mu\sigma)^{-N/3}\int_0^{\infty}x^{N+2}
\left|\chi_{\lambda}(x)\right|^2dx.
\label{rav}
\ee

Then the difference of the masses of the two eigenstates corresponding to $j_2=\frac12$ and
$j_2=\frac32$ is
\be
M_{P_{1/2}}-M_{P_{3/2}}=-\frac32E_{so_2},
\ee
so that the picture of the splitting depends on the sign of the coefficient (\ref{Eso2}).
Numerically this difference equals to $+9MeV$ for $D$ mesons and $+11MeV$ for $B$'s.
Of course the considered limit $m_Q\to\infty$ is not realistic; 
it might be reasonably well justified for the $b$-quark, but not $c$-quark. 
In what follows the heavy-quark mass is kept finite.

\begin{table}[t]
\begin{center}
\small
\begin{tabular}{|l|c|c|c|c|c|c|c|}
\hline
               &&Our&Ref.\cite{isgur2}&Ref.\cite{Faustov}&Lat.\cite{lattice2}&PDG&CLEO\\
\hline
$1P_1$\scriptsize $(P_{3/2})$&$D_1$&2.428    &2.44               &2.414               &2.405                  &2.422         &2.425\\
\hline
$1P_0$         &$D_0$&2.43     &2.4                &2.438               &2.444                  &              &     \\
\hline
$1P_2$         &$D_2$&2.445    &2.5                &2.459               &2.445                  &2.459         &     \\
\hline
$1P_1$\scriptsize $(P_{1/2})$&$D_1$&2.469    &2.49               &2.501               &2.414                  &              &$2461^{+0.041}_{-0.034}$\\
\hline
\end{tabular}
\end{center}
\caption{Masses of the $P$-level $D$ mesons in $GeV$. See Refs.\cite{PDG}-\cite{ALEPH} for the cited experimental values.}
\end{table}

\begin{table}[t]
\begin{center}
\small
\begin{tabular}{|l|c|c|c|c|c|c|c|c|c|c|}
\hline
               &&Our&Ref.\cite{isgur2}&Ref.\cite{Faustov}&Lat.\cite{lattice2}&OPAL&L3&DELPHI&CDF&ALEPH\\
\hline
$1P_1$\scriptsize $(P_{3/2})$&$B_1$&5.716    &                   &5.719               &5.684                  &     &5.67 &      &5.71&     \\
\hline
$1P_0$         &$B_0$&5.722    &5.76               &5.738               &5.754                  &5.839&     &      &    &     \\
\hline
$1P_2$         &$B_2$&5.724    &5.8                &5.733               &5.77                   &     &5.768&5.732 &    &5.739\\
\hline
$1P_1$\scriptsize $(P_{1/2})$&$B_1$&5.741    &                   &5.757               &5.73                   &5.738&     &      &    &     \\
\hline
\end{tabular}
\end{center}
\caption{Masses of the $P$-level $B$ mesons in $GeV$. See Refs.\cite{PDG}-\cite{ALEPH} for the cited experimental values.}
\end{table}

From Tables IV,V one can see that the predictions of our method for the masses of the 
$1P_1$ states are in good agreement with the lattice calculations \cite{lattice2} as well as 
with the experimental data. Namely, as far as the spectrum of $D$ mesons is concerned, 
we have good coincidence with the results of CLEO \cite{CLEO} (see Table IV). In the
$B$-mesonic spectrum we identify the state $B_1$ with the mass $m(B_1)=5.71\pm 0.02GeV$, 
recently claimed by CDF \cite{CDF}, with the lightest member of the $J=1$ doublet, 
whereas the heaviest one can be associated with the resonance reported by OPAL \cite{OPAL} (see Table V).
Unfortunately, experimental resolution does not enable one to disentangle both $P_1$
states simultaneously, so that at present time one rather has to rely on available lattice 
simulations. In such a situation other models, as well as improved lattice calculations,
are welcome to
attack this problem to have well established and clear predictions for experimentalists.

\begin{figure}[t]
\centerline{\epsfig{file=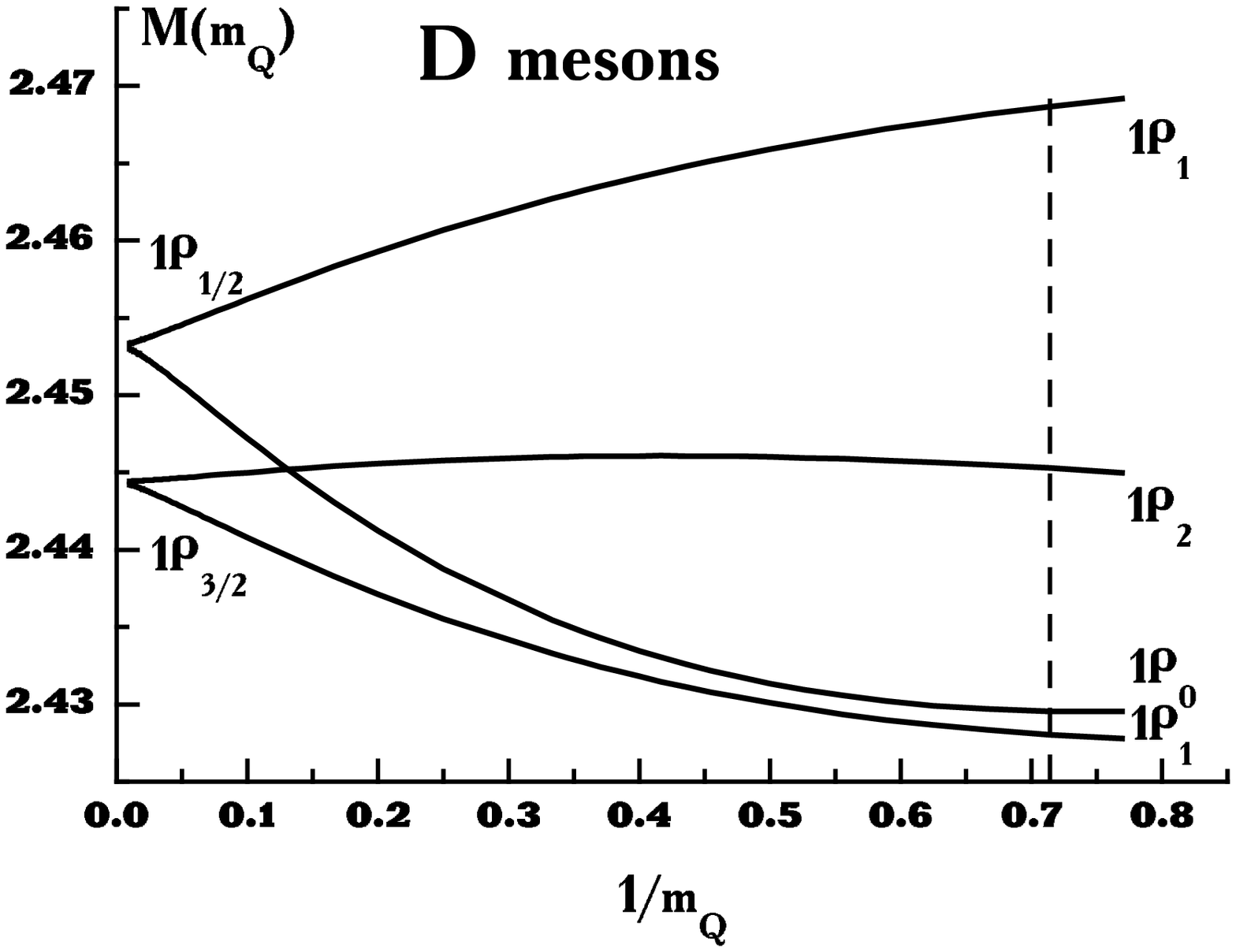,width=9.5cm}\hspace*{-1cm}\epsfig{file=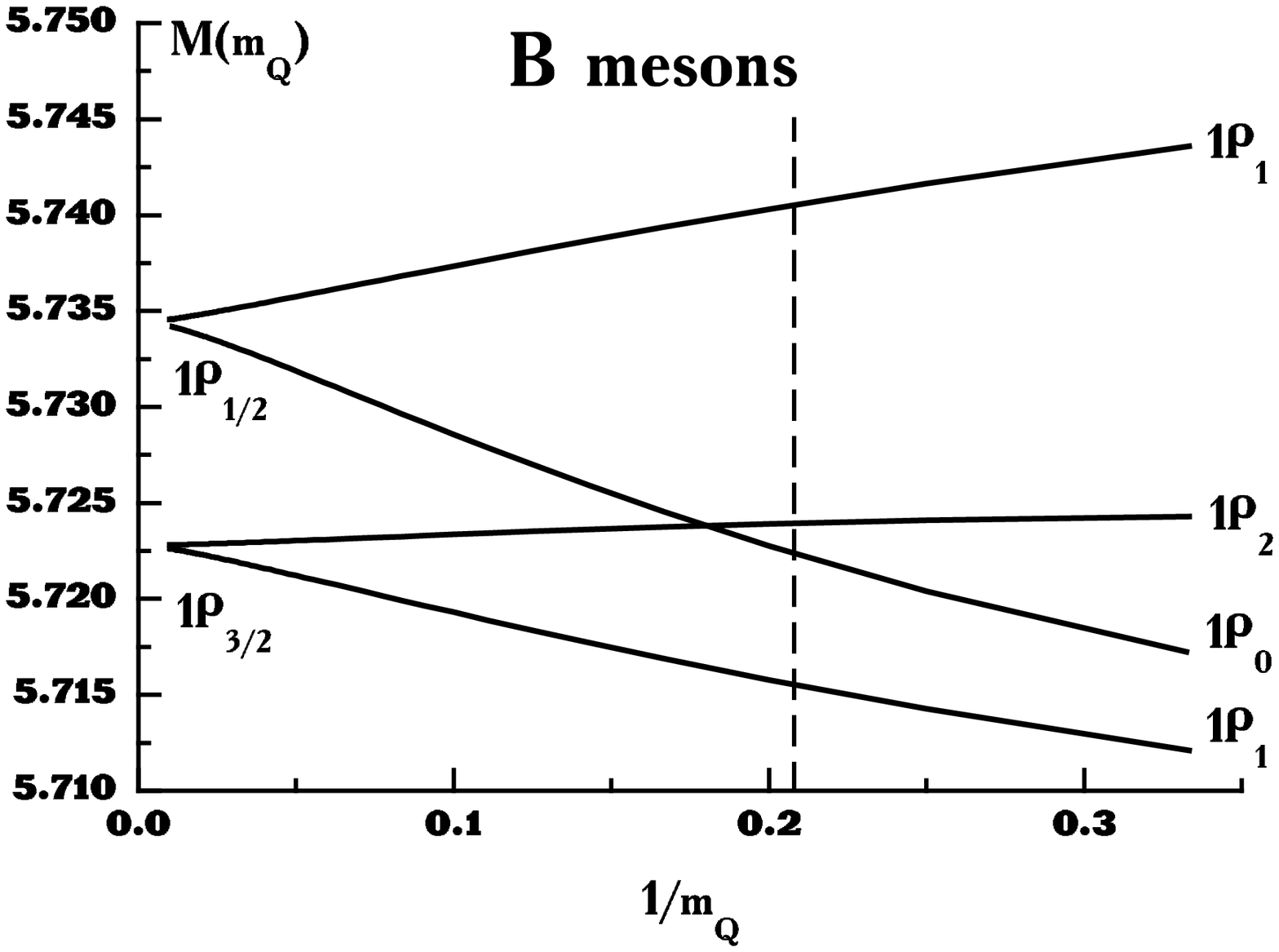,width=9.5cm}}
\caption{Splitting pattern for the $D$- (left plot) and $B$-mesonic (right plot) $P$ levels as a
function of the heavy-quark mass. The vertical dashed line corresponds to $m_Q=m_c=1.4GeV$
for the left plot and $m_Q=m_B=4.8GeV$ for the right one, respectively.}
\end{figure}

In Fig. 1 we give the splitting pattern for the $1P$ levels for both, $D$ and $B$, mesons 
for the
heavy-quark mass varying from infinity to about $1.3GeV$ with the vertical dashed line
giving the actual masses of $c$ and $b$ quarks for $D$ and $B$ mesons, respectively. 
It is also worth mentioning that according to our model the $1P_0$ and $1P_2$
levels change their ordering around the heavy-quark mass $m_Q\approx 7.9GeV$ for the
$D$-like meson (left plot in Fig. 1), and around $m_Q\approx 5.5GeV$ for the $B$-like one 
(right plot in Fig. 1).

In conclusion, let us briefly summarize the results reported in this letter. We addressed 
the question on
masses and splitting pattern of the $P$-level $D$ and $B$ mesons in the method of the QCD
string with quarks at the ends. We took into account the proper dynamics of the QCD
string,
encoded in the so-called string correction, and supplied the interquark interaction with
the one-loop corrections adapted to the case of the self-consistently generated dynamical masses of
the quarks. Using the standard values for the string tension, the strong coupling constant
and the current quark masses, we calculated the spectrum of $P$-level $D$ and
$B$ mesons and found good agreement of our results with the lattice and experimental data,
including those reported recently. Finally, we give our predictions for the
splittings between the $P$ states.
\bigskip

The authors are grateful to A. M. Badalian and Yu. A. Simonov for drawing their attention to
the recent experimental paper \cite{CDF} and stimulating discussions.
One of the authors (A. N.) would like to thank the
staff of the Centro de F\'\i sica das Interac\c c\~oes Fundamentais (CFIF-IST) for cordial
hospitality during his stay in Lisbon. The financial support of RFFI grants
00-02-17836 and 00-15-96786, INTAS-RFFI grant IR-97-232, and INTAS CALL 2000-110 is 
acknowledged. One of the authors (A. N.) is also supported via RFFI grant 01-02-06273.

\end{document}